\documentclass[11pt,preprint]{aastex}

\def\icarus{Icarus}
\usepackage{natbib}
\slugcomment{AJ, in press}

\begin{document}

\title{Water ice in the Kuiper belt}
\author{M.E. Brown}
\affil{Division of Geological and Planetary Sciences, California Institute
of Technology, Pasadena, CA 91125}
\email{mbrown@caltech.edu}
\author{E.L. Schaller}
\affil{NASA Dryden Aircraft Operations Facility, Palmdale, CA 93550
and
National Suborbital Education and Research Center, University of North
Dakota, Grand Forks, ND, 85202}
\author{W.C. Fraser}
\affil{Division of Geological and Planetary Sciences, California Institute
of Technology, Pasadena, CA 91125}

\begin{abstract}
We examine a large collection of low resolution near-infrared spectra
of Kuiper belt objects and centaurs in an attempt to understand the
presence of water ice in the Kuiper belt. We find that water ice on
the surface of these objects occurs in three separate manners: (1) 
Haumea family members uniquely show surfaces of nearly pure water ice,
presumably a consequence of the fragmentation of the icy mantle of 
a larger differentiated proto-Haumea; (2) large objects with absolute
magnitudes of $H<3$ (and a limited number to $H=4.5$) have surface
coverings of water ice -- perhaps mixed with ammonia -- that appears
to be related to possibly ancient cryovolcanism on these large objects; and
(3) smaller KBOs and centaurs which are neither Haumea family members
nor cold-classical KBOs appear to divide into two families (which we 
refer to as ``neutral'' and ``red''), each of
which is a mixture of a common nearly-neutral component and either a 
slightly red or very red component that also includes water ice. A model
suggesting that the difference between neutral and red objects
s due to formation
in an early compact solar system either inside or outside, respectively,
 of the $\sim$20 AU
methanol evaporation line is supported by the observation that methanol
is only detected on the reddest objects, which are those
 which would be expected to have
the most of the methanol containing mixture.
\end{abstract}


\section{Introduction}
The Kuiper belt is composed of low-temperature remnants of the outer
regions of the protoplanetary disk which never became incorporated 
into planets. By analogy to short-period comets, which are derived
from the Kuiper belt, and from cosmochemical considerations, it is expected
that water ice is a major constituent of the composition of Kuiper belt
objects (KBOs). That water ice was the first clearly identified
constituent on the surfaces of small KBOs and of centaurs (former KBOs
which are currently on short-lived planet-crossing orbits) was thus
not a surprise \citep{1998Sci...280.1430B, 1999ApJ...519L.101B}. 

For a large majority of KBOs which have been studied, water ice 
remains the only identifiable surface constituent, even though 
the detectable absorption features are so small that, in most cases, 
it is clear that water ice is a relatively minor component of the
surface \citep{2006ApJ...640L..87B, 2009Icar..201..272G}. One major exception is Haumea and its
satellites and collisional family, which appear to have surfaces
composed of nearly pure water ice, thought to be exposed when the 
differentiated icy mantle of the proto-Haumea was removed in a giant
impact  \citep{2007Natur.446..294B}. The other major exception is the largest Kuiper belt
objects, which are cold and massive enough to maintain volatile atmospheres
and frosts over the age of the solar system \cite{2007ApJ...659L..61S}, and whose bedrock is
covered and thus unobservable.
For the majority of KBOs with water ice present in the spectrum,
little connection has been made between the water ice visible on the
surface and any other properties of the KBOs. No simple correlation
appears between dynamical or color properties and water ice absorption
\citep{2006ApJ...640L..87B, 2009Icar..201..272G}, so emphasis has instead mostly been on
detailed modeling to determine surface constituents \citep[i.e.][]{2011Icar..214..297B} and  quantify the absence or presence of ice.

Here we examine the moderate-sized and smaller KBOs and examine
water and other ices in detail.
The goal is to examine these objects as a class,
rather than perform detailed modeling of individual objects,
in the hope of statistically understanding the causes and states of
ices on these objects.

\section{Observations and analysis}
In order to examine the properties of ices in the Kuiper belt, we 
assemble a nearly uniform set of low-resolution ($\lambda/\Delta\lambda \sim 160$) 1.5-2.4 $\mu$m reflectance spectra, almost all
obtained using NIRC -- 
the first-generation near-infrared spectrograph at the Keck 1 telescope
\citep{1994ExA.....3...77M}.
Spectra were collected from the surveys of \citet{2000AJ....119..977B} and
\citet{2008AJ....135...55B}.
We also 
include the NIRC spectrum of Charon 
\citep{2000Sci...287..107B} obtained with the identical instrument,
and the only available spectrum of 2007 OR10, obtained with the
FIRE spectrograph at the Magellan telescope \citep{2011ApJ...738L..26B}.

To augment this existing sample, we obtained new 
spectra of 15 KBOs and centaurs with the 
NIRC spectrograph
on the Keck telescope until the spectrograph was retired in 2009. 
The observations 
and data reduction were performed identically to the 
original Keck surveys, and details
are given in Table 1. The absolute magnitudes shown in the Table are taken
from the Minor Planet Center, which is the only compilation which includes
all of the objects in our sample. Such absolute magnitudes appear to 
be systematically biased compared 
to well-measured samples \citep{2005Icar..179..523R}, so, for relative
consistency, we use only the MPC magnitudes even though more accurate 
measurements are available for some of the objects. Random uncertainties in
absolute magnitude are not reported by the Minor Planet Center but, by
comparison to \citet{2005Icar..179..523R} appear to be less than 0.3 magnitudes.
For objects with observations at multiple epochs we
took the average of the individual spectra. The KBO 19521 Chaos (1998 WH24) appeared unusual in the
previous survey, so it was reobserved to determine if it has
a unique spectral type or if the observations were faulty. 
The new spectrum closely resembles other KBO spectra, so we
assume the previous spectrum was in error and only retain the
new spectrum.
Figure 1 shows all of the newly obtained spectra.
\begin{figure}
\plotone{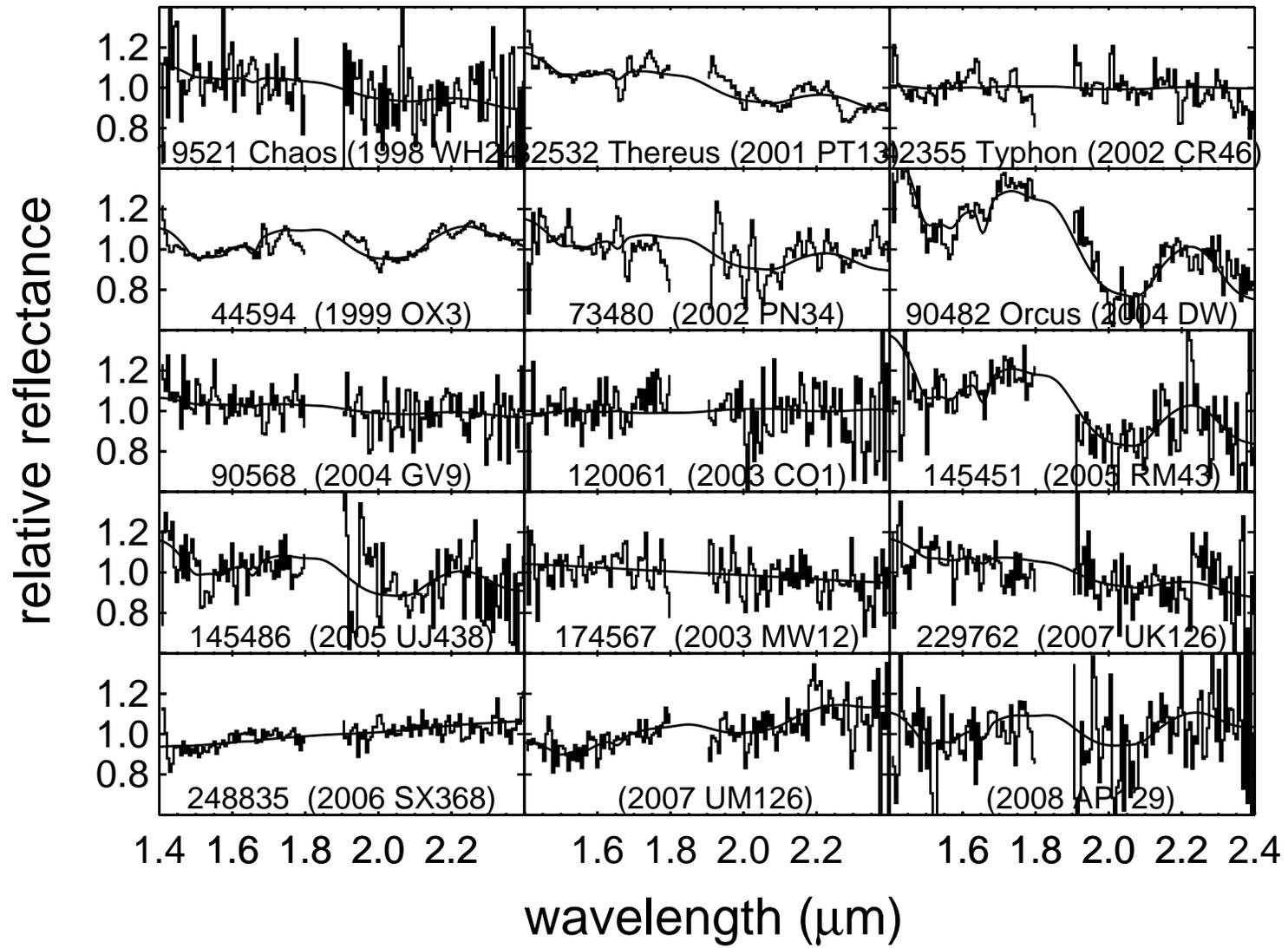}
\caption{Newly acquired spectra of KBOs and centaurs. The best fit water ice plus sloped continuum model is shown for each spectrum. All spectra
are normalized to a median value of one across the observed wavelength range.}
\end{figure}
\begin{deluxetable}{lcccc}
\tablecaption{Log of observations}
\tablehead{\colhead{object} & \colhead{date} & \colhead{exp. time} & \colhead{airmass} & \colhead{calibrator} \\ 
\colhead{} & \colhead{(UT)} & \colhead{(sec)} & \colhead{} & \colhead{} } 

\startdata
19521 Chaos (1998 WH24) & 23 Sep 2007 & 4000 & 1.02 - 1.00 & HD283798 \\
145451 (2005 RM43)  & 24 Sep 2007 & 7000 & 1.15 - 1.06 &  U Peg  \\
248835 (2006 SX368) & 24 Sep 2007 & 3400 & 1.02 - 1.04 & V Aql \\
	   & 20 May 2008 & 1200 & 1.54 - 1.06 & HD183186 \\
	   & 21 May 2008 &  800 & 1.12 - 1.02 & HD183186 \\
	   & 20 Oct 2008 & 1000 & 1.00 - 1.01 & HD211839 \\
	   & 21 Oct 2008 & 1000 & 1.02 - 1.01 & G 18 5 \\
73480 (2002 PN34)  & 24 Sep 2007 & 6000 & 1.05 - 1.07 & (V * 00 Aql)  \\
44594 (1999 OX3)   & 25 Sep 2007 & 2600 & 1.18 - 1.35 & HD 203812 \\
	   & 22 May 2008 &  400 & 1.51 - 1.38 & HD209251 \\
145486 (2005 UJ438) & 25 Sep 2007 & 1200 & 1.09 - 1.05 & LHS1365 \\
120061 (2003 CO1)   & 22 May 2008 &  800 & 1.08 - 1.50 & HD154086, HD133011, HD 132724 \\
174567 (2003 MW12)  & 22 May 2008 &  800 & 1.14 - 1.30 & HD133011 \\
	   & 06 May 2009 & 6000 & 1.10 - 1.34 & HD170363, BD-02 4024 \\
32532 Thereus (2001 PT13)  & 17 Oct 2008 & 3400 & 1.11 - 1.04 & G 18 -5 \\
	   & 18 Oct 2008 & 4000 & 1.01 - 1.07 & HD19061 \\
(2007 UM126) & 17 Oct 2008 & 6400 & 1.18 - 1.04 & LHS 1231, HD 27834 \\
	   & 19 Oct 2008 & 4000 & 1.17 - 1.08 & HD 27834 \\
229762 (2007 UK126) & 19 Oct 2008 & 2000 & 1.14 - 1.22 & HD19061, HD233399\\
           & 30 Sep 2009 & 8000 & 1.09 - 1.14 & HD289960, HD287530 \\
90482 Orcus (2004DW) 	   & 06 May 2009 & 3000 & 1.14 - 1.16 & HD95259 \\
	   & 08 May 2009 & 5000 & 1.11 - 1.22 & HD101731 \\
90568 (2004 GV9)   & 06 May 2009 & 5000 & 1.61 - 1.49 & HD136122  \\
42355 Typhon (2002 CR46)  & 07 May 2009 & 5000 & 1.05 - 1.25 & BD +00 3226 \\
	   & 09 May 2009 & 6000 & 1.05 - 1.07 & BD + 00 3226 \\
2008 AP129 & 19 Oct 2008 & 1000 & 1.34 - 1.28 & BD +45-1586\\
           & 01 Oct 2009 & 9000 & 1.12 - 1.15 & HD289960, HD287530 \\
\enddata

\end{deluxetable}

The full spectral sample (see Table 2) includes 64 objects.
To examine ices on the smaller objects, 
we remove from consideration objects known to be part of Haumea's
collisional family \citep{2007Natur.446..294B} (Haumea, 1995 SM55, 2002 TX300, 2003 OP32, 2005 RR43) and also the largest objects for which methane 
dominates the spectrum (Eris, Pluto, Makemake, and Sedna). A total
of 57 objects remain in the sample at this point.
\begin{deluxetable}{lcccccccc}
\tablecaption{Characteristics of observed KBOs and centaurs}
\tablehead{\colhead{object} & \colhead{a} & \colhead{e} & \colhead{i} & \colhead{H}&  \colhead{$f_{\rm water}$} & \colhead{uncertainty} & \colhead{$m_{\rm cont}$} & \colhead{uncertainty} \\ 
\colhead{} & \colhead{(AU)} & \colhead{} & \colhead{(deg)} & \colhead{(mag)} }
\startdata
2060 Chiron (1977 UB)& 13.7&0.38& 6.9& 6.3&-0.00&0.01& 0.08&0.00\\
5145 Pholus (1992 AD)& 20.3&0.57&24.7& 7.1& 0.13&0.01&-0.12&0.00\\
8405 Asbolus (1995 GO)& 18.1&0.62&17.6& 9.0&-0.00&0.04&-0.00&0.00\\
10199 Chariklo (1997 CU26)& 15.7&0.17&23.4& 6.6& 0.19&0.01& 0.04&0.00\\
15875  (1996 TP66)& 39.7&0.34& 5.7& 6.9& 0.00&0.08&-0.21&0.05\\
19521 Chaos (1998 WH24)& 46.0&0.11&12.0& 4.8& 0.04&0.08&-0.09&0.03\\
20000 Varuna (2000 WR106)& 43.0&0.06&17.2& 3.6& 0.01&0.02&-0.09&0.00\\
24835  (1995 SM55)& 42.0&0.11&27.0& 4.8& 0.83&0.08& 0.17&0.24\\
26181  (1996 GQ21)& 92.4&0.59&13.4& 5.2& 0.07&0.04&-0.10&0.03\\
26375  (1999 DE9)& 55.2&0.42& 7.6& 5.1& 0.10&0.05&-0.18&0.03\\
28978 Ixion (2001 KX76)& 39.5&0.25&19.7& 3.3& 0.04&0.05&-0.05&0.03\\
29981  (1999 TD10)& 99.4&0.88& 6.0& 8.7& 0.06&0.02&-0.07&0.00\\
31824 Elatus (1999 UG5)& 11.8&0.38& 5.2&10.1& 0.03&0.02& 0.06&0.00\\
32532 Thereus (2001 PT13)& 10.7&0.20&20.3& 9.0& 0.09&0.06& 0.24&0.03\\
33340  (1998 VG44)& 39.5&0.26& 3.0& 6.5& 0.06&0.18&-0.14&0.10\\
38628 Huya (2000 EB173)& 39.3&0.27&15.5& 4.9& 0.08&0.02&-0.09&0.00\\
42301  (2001 UR163)& 51.9&0.28& 0.8& 4.2&-0.01&0.13& 0.01&0.05\\
42355 Typhon (2002 CR46)& 37.7&0.54& 2.4& 7.5& 0.31&0.17& 0.05&0.10\\
44594  (1999 OX3)& 32.5&0.46& 2.6& 7.4& 0.31&0.15& 0.21&0.16\\
47171  (1999 TC36)& 39.7&0.23& 8.4& 4.9& 0.08&0.04&-0.08&0.00\\
47932  (2000 GN171)& 39.2&0.28&10.8& 6.0&-0.05&0.08& 0.04&0.03\\
50000 Quaoar (2002 LM60)& 43.3&0.04& 8.0& 2.6& 0.29&0.01& 0.07&0.00\\
52872 Okyrhoe (1998 SG35)&  8.3&0.31&15.7&10.9& 0.08&0.05&-0.02&0.03\\
54598 Bienor (2000 QC243)& 16.6&0.20&20.7& 7.5& 0.07&0.07&-0.00&0.03\\
55565  (2002 AW197)& 47.1&0.13&24.4& 3.4& 0.04&0.03& 0.03&0.00\\
55636  (2002 TX300)& 43.5&0.12&25.8& 3.2& 0.91&0.01&-1.47&0.24\\
55637  (2002 UX25)& 42.9&0.14&19.4& 3.7& 0.03&0.04&-0.02&0.00\\
55638  (2002 VE95)& 39.6&0.29&16.3& 5.6& 0.14&0.02& 0.05&0.00\\
65489 Ceto (2003 FX128)& 99.8&0.82&22.3& 6.3& 0.17&0.06&-0.10&0.03\\
66652 Borasisi (1999 RZ253)& 44.0&0.09& 0.6& 5.9&-0.22&0.17&-0.16&0.05\\
73480  (2002 PN34)& 31.2&0.57&16.6& 8.6& 0.07&0.06& 0.03&0.03\\
83982 Crantor (2002 GO9)& 19.4&0.28&12.8& 8.8& 0.16&0.05&-0.07&0.03\\
84522  (2002 TC302)& 55.7&0.30&35.0& 3.8& 0.13&0.08& 0.07&0.03\\
84922  (2003 VS2)& 39.6&0.08&14.8& 4.1& 0.09&0.02&-0.03&0.00\\
90482 Orcus (2004 DW)& 39.2&0.23&20.6& 2.3& 0.44&0.01&-0.36&0.03\\
90568  (2004 GV9)& 41.8&0.07&22.0& 4.0& 0.05&0.05&-0.04&0.03\\
95626  (2002 GZ32)& 23.0&0.22&15.0& 7.0& 0.01&0.03& 0.05&0.00\\
119951  (2002 KX14)& 38.7&0.05& 0.4& 4.4&-0.14&0.22&-0.08&0.06\\
120061  (2003 CO1)& 20.7&0.47&19.8& 8.9&-0.01&0.02& 0.08&0.00\\
120132  (2003 FY128)& 49.2&0.25&11.8& 4.9& 0.13&0.28&-0.11&0.17\\
120178  (2003 OP32)& 43.3&0.10&27.1& 3.6& 1.01&0.00&10.18&0.24\\
120348  (2004 TY364)& 39.1&0.06&24.8& 4.5& 0.06&0.02& 0.04&0.00\\
127546  (2002 XU93)& 66.6&0.69&77.9& 8.0& 0.07&0.23& 0.09&0.10\\
134860  (2000 OJ67)& 43.0&0.02& 1.1& 6.1& 0.80&0.33&-0.84&0.24\\
136108 Haumea (2003 EL61)& 43.0&0.20&28.2& 0.2& 0.66&0.00&-0.40&0.00\\
136472 Makemake (2005 FY9)& 45.4&0.16&29.0&-0.4&-1.68&0.05&-0.22&0.00\\
145451  (2005 RM43)& 92.2&0.62&28.7& 4.4& 0.32&0.05&-0.20&0.03\\
145452  (2005 RN43)& 41.7&0.03&19.2& 3.9&-0.06&0.03&-0.01&0.00\\
145453  (2005 RR43)& 43.5&0.14&28.5& 4.0& 1.01&0.00& 5.61&0.24\\
145486  (2005 UJ438)& 17.7&0.53& 3.8&10.7& 0.02&0.09& 0.03&0.03\\
174567  (2003 MW12)& 45.7&0.14&21.5& 3.4&-0.00&0.05&-0.05&0.01\\
202421  (2005 UQ513)& 43.5&0.14&25.7& 3.4& 0.09&0.04& 0.03&0.03\\
208996  (2003 AZ84)& 39.5&0.18&13.5& 3.6& 0.23&0.06&-0.10&0.03\\
225088  (2007 OR10)& 67.1&0.50&30.7& 1.7& 0.34&0.07& 0.06&0.05\\
229762  (2007 UK126)& 74.2&0.49&23.4& 3.4& 0.01&0.03&-0.11&0.03\\
248835  (2006 SX368)& 22.3&0.46&36.3& 9.5& 0.10&0.03& 0.13&0.03\\
  (2000 PE30)& 54.5&0.34&18.4& 6.1& 0.56&0.52& 0.51&0.24\\
  (2004 NT33)& 43.6&0.15&31.2& 4.4& 0.07&0.07& 0.00&0.03\\
  (2004 PG115)& 91.0&0.60&16.3& 4.9& 0.07&0.03& 0.07&0.00\\
  (2005 QU182)&114.0&0.67&14.0& 3.4&-0.17&0.05& 0.06&0.00\\
  (2007 UM126)& 12.9&0.34&41.7&10.1&-0.01&0.06& 0.20&0.03\\
  (2008 AP129)& 41.8&0.14&27.4& 4.5& 0.14&0.09& 0.04&0.05\\
  (2008 QD4)&  8.4&0.35&42.0&11.3& 0.03&0.01& 0.07&0.00\\
Charon& 39.4&0.25&17.1& 0.9& 0.74&0.03&-0.38&0.09
\enddata
\end{deluxetable}

We are interested in characterizing the amount of water ice absorption
present in each of the spectra, but
spectral models give non-unique results for the fraction
of water ice present on a surface. Differences in assumed 
grain size, type of mixing, and non-water ice components can
lead to order-of-magnitude or greater variations in the 
derived water ice abundance. Frequently, a measurement of
the depth of the 2$\mu$m water absorption feature is used as
a model-independent proxy for the amount of water ice present
\citep{2000AJ....119..977B, 2001AJ....122.2099J, 2011Icar..214..297B}, but this method ignores much of the
information contained in a broader spectrum.
\citet{2008AJ....135...55B}, therefore,
developed a simple parametric method to fit a spectrum. While the 
derived parameters cannot be directly converted to a composition,
they uniquely describe the basic characteristics of 
the spectrum of the object and thus
facilitate quantitative comparison. We use a modified 
version of that parameterization here.

We assume that in the region from 1.4 to 2.4 $\mu$m the spectra
of these KBOs can be modeled to the accuracy of the currently available
data by a simple model consisting of a linear mixture of 
a water ice spectrum and a sloped 
continuum. In our parameterization, the model spectrum, $s[\lambda]$,
 is simply given by
\begin{equation}
s[\lambda]=f_{\rm water}\ s_{\rm water}[\lambda]+(1-f_{\rm water})[m_{\rm cont}(\lambda-1.74\mu{\rm m})+.49],
\end{equation}
where $f_{\rm water}$ is a parameter which scales the amount of water ice versus 
continuum, $s_{\rm water}$ is a modeled water ice spectrum, $m_{\rm cont}$ is the slope
of the continuum added to the water ice spectrum, and $\lambda$ is
the wavelength in $\mu$m. For a modeled water ice spectrum we used
optical constants from \citet{1998JGR...10325809G} at temperatures
of 50 K and created a model spectrum assuming grain sizes of 50 $\mu$m
using the method of \citet{1993tres.book.....H}. Our goal is to have a single 
representative water ice spectrum to use in our parameterized fit
rather than to fit for additional parameters of the water ice spectrum
such as grain size or temperature. The sloped continuum is constrained 
to have a reflectance of 0.49 at 1.74 $\mu$m 
to match that of the water ice component that we used. 
Without a direct measurement of 
the true albedo of the object, however, the actual albedo of the
continuum component is unconstrained, but the parameter $f_{\rm water}$
approximates how much of the spectrum is being modeled by
water ice versus how much by continuum. We will use the term 
"fraction of water ice the in the spectrum" to describe this parameter,
but we note again that there is no unique way to convert $f_{\rm water}$ into
a fraction of water ice on the surface.

We estimate the uncertainties in the spectra
by median smoothing each spectrum with an 11 pixel wide box and then
calculating the standard deviation of the difference between the original
spectrum and the smoothed spectrum. We then perform a $\chi^2$ 
minimization to find the best-fit values for the two parameters, $f_{\rm water}$,
the fraction of water ice in the spectrum, and $m_{\rm cont}$ the slope of the continuum
using only data between 1.45 and 1.8 $\mu$m and 1.95 and 2.3 $\mu$m, where
atmospheric transmission and spectroscopic throughput are highest.
After minimizing, we create a grid of $f_{\rm water}$ and $m_{\rm cont}$ and find  $\chi^2$
for each point. Formally, the $1 \sigma$ errors on the parametric
fitting are where $\chi^2$ increases from its minimum to 1 above the
minimum. We find, however, that, by eye, 
these uncertainties appear unreasonably small,
which is not unexpected. Error bars derived from a $\chi^2$ minimization
will only give correct results if the model perfectly describes the
data and the error bars on the data are perfectly gaussian. Neither
of these conditions is likely to be true. 
After extensive experimentation with fitting by hand and examining the
$\chi^2$ values in spectra where we believe we do and do not credibly
detect water ice, we have adopted errors on our parameters to include
the regions where $\chi^2$ increases to 5 above the minimum. 
While it is difficult to assign a rigorous statistical meaning to
the uncertainties, we believe them to be the best representation of
credible uncertainties.
Full fit parameters of all objects in our sample are given in Table 2, 
and the fits to the new spectra are shown in Figure 1. 

\section{The mid-sized Kuiper belt objects}
Of the 57 objects in the spectral survey, most have uncertainties
in $f_{\rm water}$  clustered below 0.1, but 10 have significantly
larger errors. We deem those spectra too noisy for reliable analysis
and discard those 10 objects from further consideration.
In Figure 2, we show the
fraction of water ice in the spectrum as a function of absolute
magnitude for the remaining 47 objects of the sample.
A trend of higher water ice absorption on the intrinsically brightest
 objects is clearly present.
Below an absolute magnitude of $H=3$, all KBOs have significant
water ice absorption, which generally increases with lower absolute
magnitude. Above an absolute magnitude of $H=4.5$, 
no trend with size is apparent. A similar result is shown
in \citet{2011Icar..214..297B}.
While such a correlation of absolute magnitude and presence of
water ice might be expected simply from the increased albedo
of objects with more water ice on their surfaces, Spitzer radiometry
has shown that, discounting the Haumea family members, 
the objects with the largest water ice absorption
are indeed the largest objects, and not just the objects with highest 
albedos \citep{2008ssbn.book..161S}.
\begin{figure}
\plotone{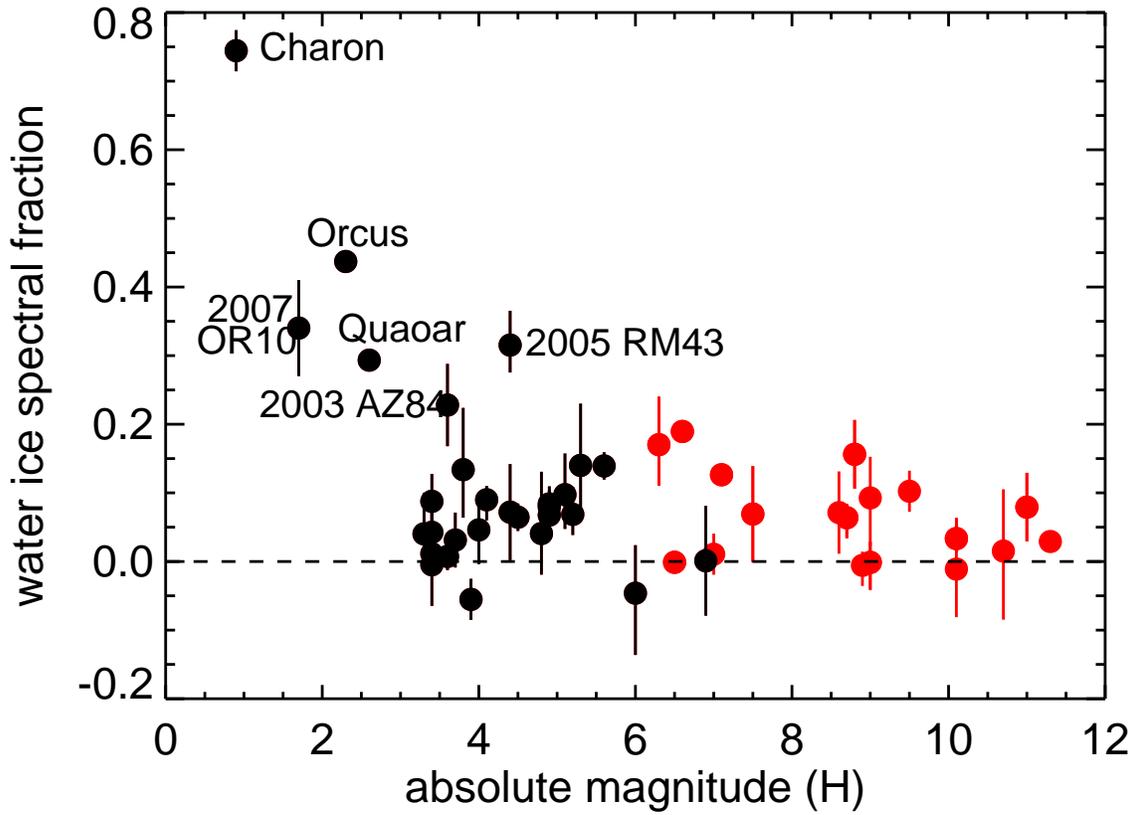}
\caption{Water ice absorption as a function of absolute magnitude. Objects with semimajor axis outside of 30 AU are shown in black,
while those inside of 30 AU are shown in red. Absolute magnitudes are taken 
from the Minor Planet Center and have typical random errors below 0.3 magnitudes}
\end{figure}

The trend of increasing surface water ice fraction with increasing size for
the largest objects is in marked contrast to the trend
of increasing density -- and thus decreasing {\it bulk} water ice
fraction -- with increasing size \citet{myannualreviews}. 

\citet{2009Icar..202..694D} and \citet{2010A&A...520A..40D} considered the possibility of surface
water flow on Charon and on Orcus -- two of the objects with the largest
fraction of water ice on their surfaces -- and concluded 
that liquid water interiors could have been present in the past
and resulted in water or water-ammonia flows through cracks to
the surface. 
The detection of ammonia on Charon \citep{2000Sci...287..107B, 2001AJ....121.1163D,2007ApJ...663.1406C,2010Icar..210..930M} and likely
also on Orcus \citep{2005A&A...437.1115D, 2008A&A...479L..13B,  2011A&A...534A.115C} supports this idea.
While modeling has yet to explore much parameter space,
it seems clear that this type of cryovolcanic activity and the amount
of water emplaced on the surface would
increase with increasing size of the object. 

It is instructive to note that
simple order-of-magnitude calculations support the idea that somewhere
around a diameter of 500 km, liquid water could have been an important
component of the early object. For example, if all of the accretional
gravitational
potential energy were converted into heat, an object 
with a 1:2 ice-rock mass mix with a diameter of 700 km would
have enough energy to melt all of its ice. Similarly, assuming
initial chondritc heating rates (including no short-lived radionuclides)
for the rock fraction
\citep{2007Icar..190..179C}, an object 500 km in diameter
would have enough radioactive heating to melt all of its ice 
in only 60 kyr.
Neither of these scenarios represents a realistic assumption about 
heat flow in these bodies, but they provide interesting limits which 
point to the 500 km size as a potentially important transition size.

Simply having cryovolcanic flow on the surface in the past
does not guarantee that the
ice would still be visible today. 
In the scenarios
 of \citet{2009Icar..202..694D} and \citet{2010A&A...520A..40D},
the cryovolcanic flow occur as the liquid interior begins to freeze and 
squeeze water out through cracks, a process which would necessarily be
delayed. \citet{2009Icar..202..694D} proposes that the cryovolcanic flows on Charon
are actually recent, but \citet{2010SSRv..153..447S} suggests that
the physical parameters they require for current cryovolcanism are implausible.
From the simple existence of water ice on the surface, however, there is no need to invoke current or even recent 
cryovolcanism; the very fresh water ice on the surface of Haumea and its
collisional family \citep{2007Natur.446..294B} combined with dynamical modeling which
shows that the surfaces have likely been exposed for most of the
age of the solar system \citep{2007AJ....134.2160R, 2008AJ....136.1079L} demonstrate that fresh
water ice surfaces can be preserved on billion year time scales in
the outer solar system.

If the water ice on the surfaces of the medium-sized KBOs is indeed from
cryovolcanic flows, and if the early
interior oceans of these medium-sized KBOs were
of similar composition, we would expect that, like Charon and apparently Orcus,
 all of the objects would
also have ammonia extruded onto their surfaces with the water ice. 
Some of the larger and colder objects
retain a tiny amount of methane, whose strongest absorption feature is in
the same location as that of ammonia, so confirmation of ammonia absorption
features is
difficult. The objects smaller than Orcus currently do not have spectra
with sufficient signal-to-noise to be able to test for the presence of
ammonia.

\section{Smaller objects}
Many objects with absolute magnitude $H>3$ (corresponding 
approximately to a diameters of 800 km and smaller; Stansberry et al., 2008) are
consistent with having little or no water ice detected on their surface, 
but almost all of these objects have $f_{\rm water}>0$.
Photon noise or systematic error from telluric correction -- the two
largest sources of uncertainty in the data -- could not produce a bias towards
positive values of $f_{\rm water}$.
Indeed, we regard the nearly complete lack of 
objects which are constrained to negative values of $f_{\rm water}$ within
their uncertainties as an indication of the
robustness of our method. We conclude, therefore, that even the low
level of water ice fraction detected in the majority of the
objects is a real indication that water ice is common within 
the spectra of
the smallest measurable objects.

To date, no trends have been found or explanations proposed for the
low levels of water ice on the smaller objects. \citet{2006ApJ...640L..87B}
found no simple correlation with color or orbital parameters;
a recent analysis of a large sample of spectra \citep{2011Icar..214..297B}
similarly found no correlations.

Recently, however, 
\citet[hereafter the H/WTSOSS survey]{hwtsoss} 
have used a large visible-infrared
photometric survey from the Hubble Space Telescope to suggest that -- when the
cold classical KBOs and Haumea family members are excluded from the sample -- 
small KBOs and Centaurs consist of two distinct compositional
families. Each family is described by a mixture between a common dark neutrally colored
substance and either a slightly red (for the neutral family) or
very red (for the red family) component. Both red components
are characterized by a decrease in reflectivity between 1.38 and 1.53 $\mu$m,
which is consistent with a contribution from water ice in addition
to whatever is causing the red coloration. 
In order to analyze if such a trend is visible in
our measured water ice fraction, we examine the water ice spectral fraction
of all objects with $f_{\rm water} < 0.2$ (all KBOs with $f_{\rm water}>0.2$
appear to have surface characteristics influenced by their larger size 
thus we discard them from the sample).
as a function of color (Figure 3). We 
use optical photometry from the MBOSS compilation\footnote{
available at http://www.eso.org/~ohainaut/MBOSS/} to construct a spectral
gradient using the method of
\citet{2002A&A...389..641H}, but only using the B, V, R, and I
colors, to more closely match the H/WTSOSS measurements. The spectral
calculated spectra gradient for the full sample can be found in Table 2.
Colors have been measured for 29 of our 41 $f_{\rm water}<0.2$ objects.
Like previous analysis, we find no simple
trend between water ice absorption and color. However if we examine the neutral objects (which we empirically define as spectral
gradient less than 17\%/100 nm)
and red objects (spectral gradient greater than 17\%/100 nm) separately
we begin to see the trend which would be predicted by the H/WTSOSS survey:
within each color group, water ice absorption becomes stronger as
the objects become redder. The uncertainties in the water ice absorption
are sufficiently high that obtaining robust statistics is impossible,
though
for the red objects, a rank correlation test shows that there is a positive
correlation between the color and the fraction of water ice absorption at
the 87\% confidence level.  Measurement of the colors of the 12 objects for
which no such measurements yet exist will help to better define the 
behavior of the data.
\begin{figure}
\plotone{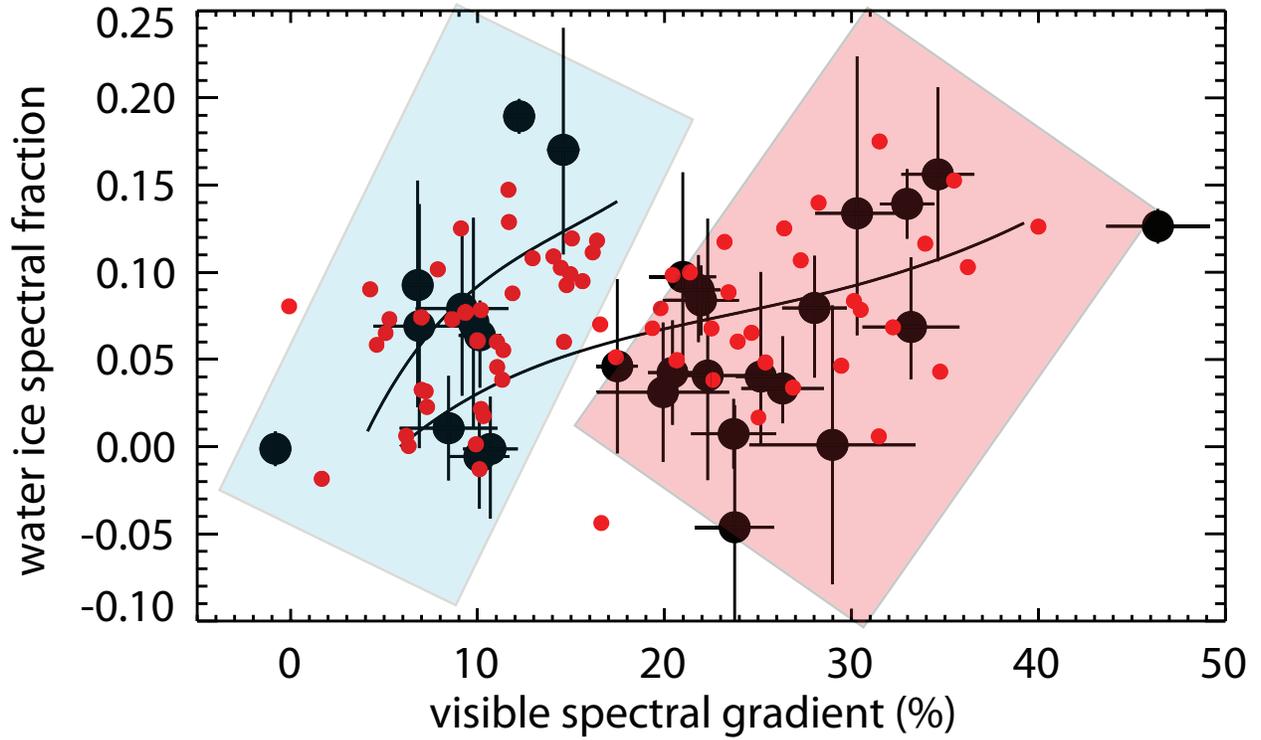}
\caption{Water ice absorption fraction as a function of spectral color gradient. Black points with error bars 
show the spectral data and colors derived from visible photometry.
The red points without error bars (typical uncertainties are smaller than those of the spectral data) are from 
the H/WTSOSS photometric survey of \citet{hwtsoss} and have been transformed to these variables. The two lines show
the mixing models which best fit the three-color H/WTSOSS data. The blue and red shading show the neutral and the 
red families of objects.}
\end{figure}

To further examine the relationship between these
spectral results and the photometric results of the H/WTSOSS survey
we derive an approximate transformation from the HST photometric colors to
our color gradient-water ice absorption system.
First, we use the 41 objects which have both ground-based spectral
gradient measurements and H/WTSOSS F606W and F814W photometry
to derive an empirical
second-order polynomial transformation from [F606W]-[F814W] color to
spectral gradient. We then model a series of water ice plus continuum 
spectra with increasing values
of $f_{\rm water}$ and we measure the equivalent
[F139M]-[F152M] H/WTSOSS color
for each of these spectra. The relationship between H/WTSOSS color
and $f_{\rm water}$ is nearly linear, but we fit a second order polynomial
to more precisely define our transformation. 

Using these transformations, we show the \citet{hwtsoss} mixing models which provide the
best fit to the H/WTSOSS three-color data. 
The results show that not only do the
mixing models qualitatively suggest that the red ends of the two color
families should have more water ice absorption, but, remarkably, the
mixing models also predict the overall magnitude of the water
ice absorption correctly. To further examine the correlation between the
H/WTSOSS and the spectroscopic data, 
we show all of the H/WTSOSS data for $H>5.5$ (this was the absolute
magnitude that \citet{hwtsoss} chose to analyze because of the
large number of both KBOs and Centaurs fainter than this value) non-cold classical and
non-Haumea family KBOs converted to our visible gradient - water absorption
system. The photometric measurements of water ice follow
the same general behavior and occupy the same range as the spectroscopic
measurements. (Note that both the photometric and spectroscopic 
results suggest that the common end-member of the mixing
models is likely redder than suggested by \citet{hwtsoss}. A possible 
reason for this discrepancy is that the end-members for the mixing
model were restricted in
color space to be within the range of measured object colors, which
appears unlikely to be true.) This remarkable correspondence between
the spectroscopic and photometric data suggests that both data
sets are viewing the same phenomenon; the H/WTSOSS data truly
do provide a measurement of water ice absorption, and the spectroscopic
data support the same dual mixing model interpretation as the H/WTSOSS
data.

More remarkably, this interpretation suggests that
we can now identify another of the components of
the mixing model. \citet{hwtsoss} suggested that the common 
end-member is consistent with the colors of hydrated silicates. 
These data suggest that the two other end-members both contain
water ice as part of their constituents.
Water ice cannot be the only constituent, however, as both of
these end-members are more red that water ice. It is clear, then,
that the water ice must be mixed with some other component in
the end-members to provide the red color. Interestingly, the water ice 
fraction in the end member must remain nearly constant even as the
end member is mixed with varying amounts of the silicate-like end member.

\section{Methanol}
Irradiated hydrocarbons appear a natural explanation for the red
materials in the two red end-members. In the evaporation gradient
hypothesis of \citet{2011ApJ...739L..60B}, surfaces in the early Kuiper belt bifurcate
into those cold enough to retain methanol and those too hot for methanol
to remain on the surface. In
this hypothesis one end member would be an irradiated  mixture of
CO$_2$ and water ice, while the other would be an irradiated mixture of water ice, CO$_2$,
with methanol added, causing a redder color \citep{2006ApJ...644..646B}. 

Methanol is the only other ice that has been suggested to be present in the spectra
of small KBOs or centaurs. An  absorption feature
around 2.27 $\mu$m twas first observed on the centaur Pholus \citep{1993Icar..102..166D}
and later suggested to be due to methanol or a similar light hydrocarbon 
\citep{1998Icar..135..389C}. Subsequent observations suggested 
that the KBOs 1998 GQ21 and 2002 VE95 had spectra very similar to that of Pholus,
including the 2.27 $\mu$m absorption \citep{2006ApJ...640L..87B, 2011Icar..214..297B}.
The objects Pholus, 1996 GQ21, and 2002 VE95 are three of the four
reddest objects in our sample.
Recently, it has been suggested that many more objects might have these
2.27 $\mu$m absorption features though the don't necessarily share
the other characteristics of the surface of Pholus (moderate water
ice absorption, very red color) \citep{2011Icar..214..297B}.

Robust detection of narrow features at this wavelength is difficult for these
faint objects. 
Nonetheless, we believe that a statistical
assessment might still discern trends in methanol absorption even
if the individual spectra have less reliable results. 
We therefore model all of the spectra allowing an additional 
methanol ice component, where the methanol spectrum is
taken from \citet{1998Icar..135..389C}. We define the methanol ice fraction,
$f_{\rm methanol}$ similarly to $f_{\rm water}$ for water ice
and derive the uncertainty in the same manner. Results for all objects
in the extended sample are listed in Table 2.
To better analyze the data we first discard
all of the measurements for which the uncertainties in $f_{\rm methanol}$ are greater
than the median uncertainty in our restricted sample. We then examine the
methanol absorption fraction as a function of object color (Figure 4).
The red family of objects (with spectral gradient greater than
17\%/100 nm) shows a positive correlation of methanol absorption 
with color at the 98.8\% confidence level. 
The blue family of
objects shows no such correlation. Indeed, the only blue object
on which methanol is potentially detected at a significant
level -- the centaur Thereus -- was
examined with much higher signal-to-noise by \citet{2005A&A...444..977M},
who strongly rule out any methanol absorption at the level suggested
here. While it is possible that the spectrum is strongly rotationally 
variable, we suspect, instead, that our measurement of methanol is
simply spurious. Further observations of this object are warranted.
\begin{figure}
\plotone{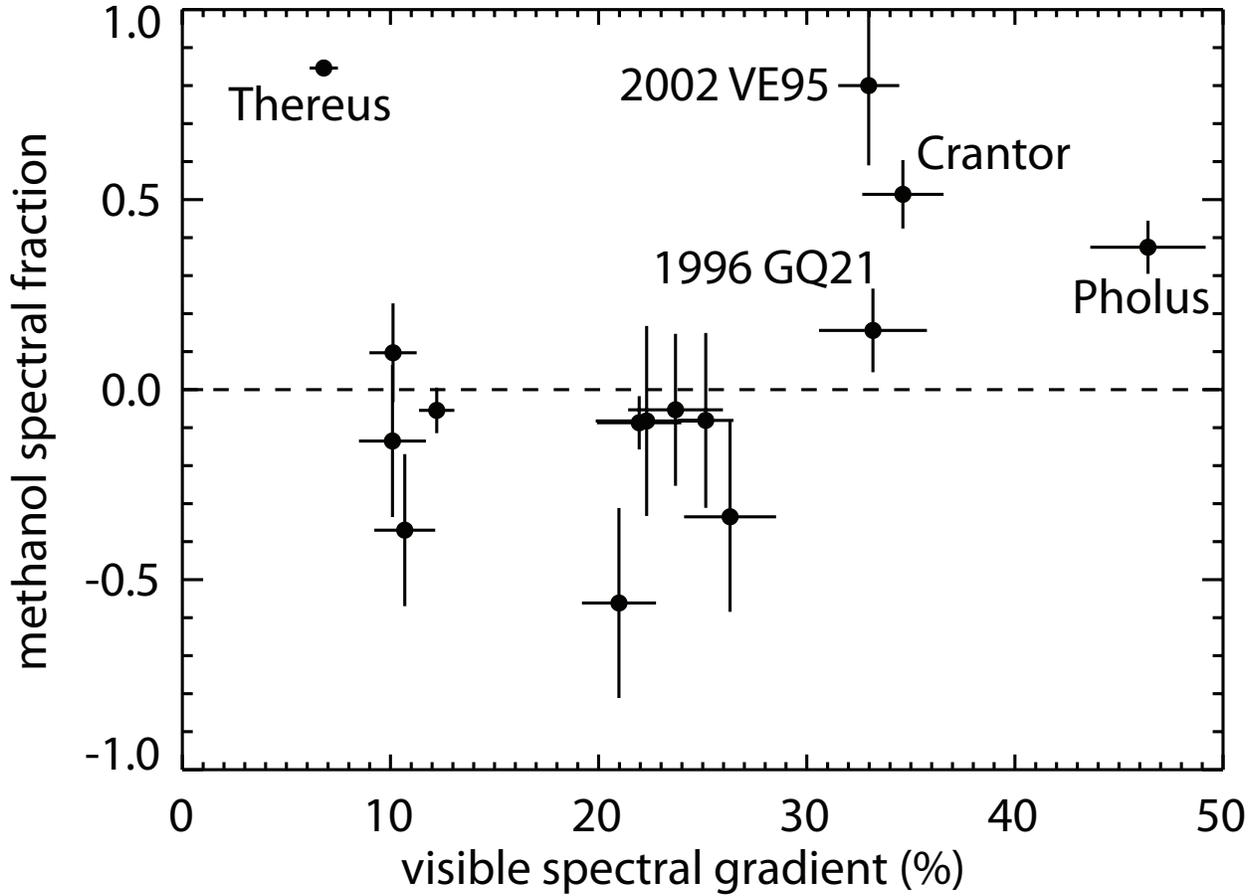}
\caption{Methanol ice absorption fraction as a function of spectral color gradient. The red family shows a positive correlation
between color and methanol absorption and color at the 99.8\% confidence level, as would be predicted by the methanol
evaporation model for KBO coloration. In the blue family, methanol is only detected on the centaur Thereus, but higher
quality spectra suggest that this detection is spurious.}
\end{figure}

Of the objects in the red population, all four for which our
method reports robust detections of
methanol have previously been identified as containing methanol-like
absorption features
-- Pholus \citep{1998Icar..135..389C}, 1996 GQ21 \citep{2011Icar..214..297B} , 2002 VE95 \citep{2006ApJ...640L..87B}, and Crantor \citep{2007A&A...475..369A}.

The appearance of methanol conforms to the expectation of the
evaporation gradient model \citet{2011ApJ...739L..60B} combined with the H/WTSOSS
mixing model. No clear detections of
methanol are made in the blue population, while in the red population 
methanol is only clearly detected on the reddest end, where the mixing
model predicts the most ices.

\section{Discussion}
The spectroscopic and photometric analysis here shows that water
ice in the Kuiper belt comes through three separate distinct processes.
Nearly pure water ice is seen only on Haumea and on Haumea's 
collisional family (and satellites) 
and must be related to the impact that created
the family. The small sizes of the family members, coupled with
the low densities of Haumea's satellites \citep{2009AJ....137.4766R} suggests
that these bodies are undifferentiated and nearly pure water ice 
throughout. Such pure water ice surfaces could be formed if 
the collisional family is fragments of a nearly pure icy mantle 
in a early differentiated proto-Haumea. Fragments from the crust of
the proto-Haumea, which would presumably show a darker hydrocarbon
irradiated surface, have not been identified, but might be present. Crust
fragments will be rare compared to mantle fragments, however.

KBOs with absolute magnitudes smaller than $H=3$ 
(diameters $\gtrsim 800 $km) have 
increasingly abundant water ice on their surfaces, presumably as a result
of cryovolcanism at some point in the (possibly quite distant) past.
Some objects between $H=3$ and $H=4.5$ also have abundant water ice,
while none with $H>4.5$ have as much. No obvious characteristics
separate the more water ice rich objects 2003 AZ83 and 2005 RM43 from
their similarly-sized less-icy objects; differences in formation location,
density, or evolutionary history are obvious candidates for these differences,
but verification will be difficult.

Smaller KBOs and centaurs show a systematic trend of water ice
and color that corresponds to the trends found in the
H/WTSOSS photometric survey. When Haumea family members and cold classical
KBOs are excluded from the population, the remaining objects
form two color families, each of which can be described by a mixing 
model. From the H/WTSOSS photometric data on small objects, 
\citep{hwtsoss} suggested
that the mixing component that causes an apparent absorption at 1.54 $\mu$m
could be water ice. The spectroscopic data on larger objects supports
this suggestion.

We now have possible identifications for two of the four major components
of the mixing model. \citet{hwtsoss} suggested that the 
component that is common to both families is consistent with many
common hydrated silicates. For the neutral family,
this component is mixed with a second component which is a mixture
of a slightly red material and water ice (thus the ``neutral'' family is,
in fact, slightly red). For the red family,
the common component is mixed with a second component which is
a mixture of a very red material and water ice.

The difference between the slightly red and very red materials 
is possibly caused by the absence or presence of methanol ice on the surface
of the object during early irradiation \citep{2011ApJ...739L..60B}. Objects which formed inside
of $\sim$20AU in a compact early solar system would have been too hot to retain
surface methanol for an amount of time long enough for methanol irradiation
to affect the coloration. Outside of 20 AU, objects are cold enough that methanol
remains present, possibly explaining the redder component of the red family.
This suggestion is supported by the observation that
methanol is only robustly detected on the very reddest members of
the red population.

\acknowledgements This research has been supported by grant
NNX09AB49G from the NASA Planetary Astronomy program. 
Some of the data presented herein were obtained at the W.M. Keck Observatory, which is operated as a scientific partnership among the California Institute of Technology, the University of California and the National Aeronautics and Space Administration. The Observatory was made possible by the generous financial support of the W.M. Keck Foundation. Support for program HST-GO-11644.01-A was provided by NASA through a grant from the Space
Telescope Science Institute, which is operated by the Association of the Universities for Research in Astronomy, Inc., under NASA contract
NAS 5-26555.

\clearpage

\end{document}